\documentclass[letterpaper]{scrartcl}
\usepackage{fullpage}
\usepackage{listings}
\usepackage{alltt}
\usepackage{multirow}
\usepackage{epsfig}
\usepackage[hyphens]{url}
\usepackage[top=1in, bottom=1.2in, left=1in, right=1in]{geometry}

\newfont{\secfntab}{ptmb8t at 10pt}

\begin{document}
\title{Practical Fine-grained Privilege Separation in Multithreaded Applications}
\author{
Jun Wang, Xi Xiong$^{\dag}$, Peng Liu \\
\textit{College of Information Science and Technology, Penn State University}\\
\textit{Amazon.com$^{\dag}$}\\
\texttt{\{junwang, pliu\}@ist.psu.edu, xixiong@amazon.com$^{\dag}$}\\
\\
\\
}
\date{
May, 2013\\
PSU-S2-13-051
}
\maketitle

\begin{abstract}
    \vspace{5pt}
    \begin{center}
        {\Large \textbf{Abstract}}
    \end{center}

    \vspace{5pt}
An inherent security limitation with the classic multithreaded programming
  model is that all the threads share the same address space and, therefore,
  are implicitly assumed to be mutually trusted.
This assumption, however, does not take into consideration of many modern
  multithreaded applications that involve multiple principals which do not
  fully trust each other.
It remains challenging to retrofit the classic multithreaded programming model
  so that the security and privilege separation in multi-principal
  applications can be resolved.

This paper proposes \textsc{Arbiter}, a run-time system and a set of security
  primitives, aimed at fine-grained and data-centric privilege separation in
  multithreaded applications.
While enforcing effective isolation among principals, \textsc{Arbiter}
  still allows flexible sharing and communication between threads so that the
  multithreaded programming paradigm can be preserved.
To realize controlled sharing in a fine-grained manner, we created a novel
  abstraction named \textsc{Arbiter} Secure Memory Segment (ASMS)
  and corresponding OS support.
Programmers express security policies by labeling data and principals
  via \textsc{Arbiter}'s API following a unified model.
We ported a widely-used, in-memory database application
  (\textsf{memcached}) to \textsc{Arbiter} system, changing only around 100 LOC.
Experiments indicate that only an average runtime overhead of 5.6\% is induced
  to this security-enhanced version of application.
\end{abstract}

\newpage

\section{Introduction}
While multithreaded programming paradigm has clear advantages over
multiprocessed programming, it implicitly assumes that all the threads inside
a process are mutually trusted. This is reflected on the fact that all the
threads inside a process run in the same address space and thus share the same
privilege to access resource, especially \emph{data}.  This assumption does
not take into consideration of the sharply increasing number of
multi-principal applications in recent years, wherein principals usually do
not fully trust each other and therefore should not have the same privilege
for data access. 

When two threads are not supposed to share any data, the problem can be
resolved by ``moving back'' to the multiprocessed programming. However, as
more and more applications are by-design expecting their principal threads to
share data and to collaborate, a fundamental security problem emerges; that
is, \textit{how to retrofit the classic multithreaded programming model so
that  ``thread-are-mutually-trusted'' assumption can be properly relaxed?}

This security problem, once solved, can even substantially benefit
single-principal applications: when two threads X and Y are providing
different though related functionalities and when X is compromised, the
attacker can no longer corrupt Y's data objects that are not shared with X.
We view this as a second motivation of this work.

A couple of existing solutions could be potentially utilized to approach this
problem.  First, program partitioning, such as OpenSSH \cite{Provos:PPE} and
Privtrans \cite{privtrans}, can realize privilege separation by splitting
privileged code and unprivileged code into separate compartments.
However, the privileged vs. unprivileged code separation policy in many cases
is very different from application-logic-based (data access)  privilege
separation policies.  For example, the policy used in our motivating example
in Section 2 simply cannot be ``expressed'' by the privileged vs. unprivileged
code ``language''.  Neither \cite{Provos:PPE} nor \cite{privtrans} can handle
the privilege separation needs of a set of threads that do not contain any
privileged code.
Second, multiprocessed programming is another resort. However, it requires the
shift of programming paradigm and thus programmers can no longer enjoy the
convenience of multithreaded programming. Specifically, complex
inter-process communication (IPC) protocols (e.g., pipes and sockets in Chrome
\cite{Chrome}) are needed and, thus, a considerable amount of effort is
required to design, implement, and verify these protocols.  In addition,
message passing is considered as less efficient than direct memory sharing
\cite{Singularity:Language}.
Third, language-based solutions, such as Jif \cite{JFlow} and Joe-E
\cite{Joe-E}, can realize information flow control and least privilege at the
granularity of program data object.  However, they need to rely on type-safe
languages like Java.  As stated in \cite{JFlow}, type checking and label
checking ``cannot be entirely disentangled, since labels are type constructors
and appear in the rules for subtyping.'' As the result, programmers have to
rewrite legacy applications not originally developed in a type-safe language.

This paper proposes \textsc{Arbiter}, a run-time system and a set of security
primitives, aimed at data-centric and fine-grained privilege separation in
multithreaded applications.
\textsc{Arbiter} has three main features.
First, \textsc{Arbiter} not only provides effective isolation among
principals, but also allows flexible sharing and communication between
threads.  The classic multithreaded programming paradigm is preserved so that
programmers no longer need to conduct manual program behavior controlling,
design inter-compartment communication protocols, or adopt a new language.
Second, application developers only need to express security properties of
data in a lightweight manner and then \textsc{Arbiter} will infer, configure,
and enforce a complete set of policies based on those properties.  Compared to
the program partitioning approach which requires a per-application security
protocol, \textsc{Arbiter} provides programmers with a unified security model
and interface.  Finally, \textsc{Arbiter} supports controlled sharing with
fine-grained access control policy at data-object level.  It focuses on
data-centric privileges and targets at explicit control of data flow inside
applications, which can be complementary to OS-level privilege separation
mechanisms.

We designed a novel framework, including a run-time system and OS support, to
achieve these features.  To effectively separate data-centric privileges, we
augment traditional multithreaded programming paradigm with
\emph{programmer-transparent} address space isolation between threads (In
contrast, address space isolation is not transparent to programmers in
multiprocessed programming).  To realize controlled data sharing, we created a
new OS memory segment abstraction named \textsc{Arbiter} Secure Memory Segment
(ASMS), which maps shared application data in multiple address spaces to the
same set of physical pages but with different permissions.  We designed a
memory allocation mechanism to achieve fine-grained privilege separation for
data objects on ASMS.  We also provide a security label model and a set of
APIs for programmers to make lightweight annotations to express security
policy.  We implemented an \textsc{Arbiter} prototype based on Linux,
including a run-time library, a reference monitor, and a set of kernel
primitives.

To evaluate retrofitting complexity and performance overhead, we ported a
widely used application, \textsf{memcached} \cite{website:memcached}, to
\textsc{Arbiter} system.  \textsf{Memcached} is a general-purpose, in-memory
object cache.  We analyzed the security requirements among the threads in
\textsf{memcached} and customized the corresponding security policies.  The
changes to the program, mostly annotations, are only around 100 lines of code
out of 20k total lines of \textsf{memcached} code.  Experimental results
indicate that only an average runtime overhead of 5.6\% is induced. We also
tested \textsc{Arbiter} with a group of microbenchmarks to further examine the
contributing factors of the overhead. 

The contributions of this work include:
\vspace{-5pt}
\begin{itemize}\itemsep0pt \parskip0pt \parsep0pt
	\setlength{\leftskip}{-10pt}
    \item A practical framework for data-centric privilege separation in
        multithreaded applications, which is mostly-transparent to
        programmers.  Not only is privilege separation ensured, the
        convenience and efficiency in conventional multithreaded programming
        paradigm are also preserved.
    \item A new memory segment abstraction and memory allocation mechanism for
        controlled sharing among multiple threads with fine-grained access
        control policy.
    \item A unified security model and interface with simple rules for secure
        cross-thread sharing.
    \item A full-featured \textsf{memcached} built on top of \textsc{Arbiter}
        with enhanced security and slight performance penalty.
\end{itemize}
\vspace{-5pt}

The rest of this paper is organized as follows. Section 2 provides a high-level 
overview of \textsc{Arbiter}'s approach. Section 3 and 4 discuss the design 
and implementation. Section 5 describes the retrofit of \textsf{memcached}. 
Section 6 shows the performance evaluation. Related work is described 
in Section 7. Section 8 discusses pros and cons of our approach and Section 9 
concludes.

\section{Overview}

This section begins with an example to show how programs can benefit
  from \textsc{Arbiter} system. 
Then Section 2.2 introduces \textsc{Arbiter} API and explains its usage.
Finally, Section 2.3 presents the security model that programmers can
  utilize to specify security policies.

\subsection{Motivating Example}
To demonstrate how data-centric privilege separation can protect
  program internal data,
consider a calendar server application which has a feature of 
  scheduling activities for multiple users. 
Suppose that the server is a multithreaded program written in C using
  \textsf{Pthreads} library and includes three types of threads.
First, \emph{worker} threads are used to handle the connection and communication 
  with each user client. 
Second, a \emph{scheduler} thread is responsible for figuring out time slots 
  available for all the participating users based on their calendars. 
Third, a \emph{dispatcher} thread is in charge of dispatching and
  synchronization.

Now suppose that Alice and Bob are going to schedule a meeting using 
  this scheduling server. 
However, security concerns (SC) arise when Alice and Bob want to protect 
  the confidentiality of their calendar information against each other (SC1).
  This is simply because, in a traditional multithreaded process, 
  threads could have direct access to the data belonging to other threads.
  If some vulnerabilities exist in worker threads and are exploited by Bob,
  he might be able to take control of the entire process and therefore access
  sensitive information of Alice. 
Moreover, Alice and Bob may also want to protect the integrity of their 
  calendar information against the scheduler thread, which might have design
  flaws or programming errors (SC2). This is usually the case if
  the scheduler employs a complex algorithm and some complicated
  data structures. 
Further more, the scheduler thread, together with Alice and Bob, may also 
  need to protect the secrecy of the results against the worker thread serving
  another user, say, Charlie (SC3). 
Table 1 provides a complete summary of access rights rooted from SC1, SC2, and
SC3 (along with the labels used to specify such policies, which will be
formally introduced in Section 2.3). 

\begin{table}[tbp]
\centering
{\small
\setlength{\tabcolsep}{1pt}
\begin{tabular}{c|c|c|c|c} \hline
& \textbf{Alice} & \textbf{Bob} & \textbf{Charlie} & \textbf{scheduler}\\
\textbf{Data} & L:\{dr\} & L:\{dr\} & L:\{\} & L:\{ar,br\}\\
& O:\{ar,aw\} & O:\{br,bw\} & O:\{cr,cw\} & O:\{ar,br,dr,dw\}\\ \hline \hline
\textbf{Alice's Cal.} & \multirow{2}{*}{\textbf{RW}} & \multirow{2}{*}{\textbf{--}} & \multirow{2}{*}{\textbf{--}} & \multirow{2}{*}{\textbf{R}}\\
L:\{dr,ar,aw\} & & & & \\ \hline
\textbf{Bob's Cal.}& \multirow{2}{*}{\textbf{--}} & \multirow{2}{*}{\textbf{RW}} & \multirow{2}{*}{\textbf{--}} & \multirow{2}{*}{\textbf{R}}\\
L:\{dr,br,bw\} & & & & \\ \hline
\textbf{Result} & \multirow{2}{*}{\textbf{R}} & \multirow{2}{*}{\textbf{R}} & \multirow{2}{*}{\textbf{--}} & \multirow{2}{*}{\textbf{RW}}\\
L:\{dr,dw\} & & & & \\ \hline
\end{tabular}
}
\caption{Security concerns in motivating example}
\vspace{-15pt}
\end{table}

These complex permissions are hard to achieve in traditional multithreaded
programming model.
With \textsc{Arbiter}, however, programmers can easily tackle such problems.
In order to make programs run on \textsc{Arbiter}, programmers
  simply need to write or port programs in mostly the same way as writing a 
  traditional multithreaded program, except for some lightweight annotations
  to denote the desired security policy. 
The following code fragments, implementing the above idea of secure activity
scheduling, illustrates the ease of adoption of \textsc{Arbiter} system. 

First, worker thread accepts user's input of calendar and stores it as
  a dynamic data object \texttt{struct cal}. The protection of user's
  calendar is through \textsc{Arbiter}'s secure memory allocation 
  \texttt{ab\_malloc}. 
Its second argument \texttt{L\_user} is a \emph{label} representing how the data is
shared or protected.  Here, the value \texttt{ur,uw} and \texttt{dr} can help
to resolve SC1 and SC2, that is, the calender is read-writable only to the
corresponding user and is read-only to the scheduler.  Other than that, nobody
can access that data (see Section 2.3 for how the label works).
Note that \texttt{ur,uw} will be instantiated to \texttt{ar,aw} for Alice and
\texttt{br,bw} for Bob at run time.

\lstset{
	language=C,
	basicstyle=\small,
	frame=lines,
	breaklines=true,
	breakatwhitespace=true,
	morekeywords={L\_user, O\_user, L\_ret, L\_sched, O\_sched}
}
{\small
\linespread{0.7}
\begin{lstlisting}
#define CAL_SIZE sizeof(struct cal)
thr_worker(argW_t *arg) {
  ... // receive input of user's calendar
  label_t L_user = {dr, ur, uw};
  calendar = (struct cal*)ab_malloc( CAL_SIZE, L_user);
  ... // send out result
}
\end{lstlisting}
}

Next, the scheduler thread receives the calendar of Alice and Bob
  from the worker threads and computes the available time slots.
Similarly, scheduler thread returns the results allocated using 
  \texttt{ab\_malloc}. 
The label \texttt{L\_ret} can help to ensure SC3, that is, the results are only
  readable to worker threads serving Alice and Bob.
{\small
\linespread{0.7}
\begin{lstlisting}
thr_scheduler(argS_t *arg) {
  ... // compute available time slots using certain algorithm
  L_ret = {dr, dw};
  ret = (struct cal*)ab_malloc(CAL_SIZE, L_ret);
  ...
}
\end{lstlisting}
}

Finally, regarding the dispatcher thread, it first dispatches two worker 
  threads with label \texttt{\{dr\}} and corresponding ownership 
  \texttt{O\_user}.
Here, \emph{ownership} means the possession of certain types of information
being protected (see Section 2.3 for details). 
After user inputs are collected, the dispatcher thread dispatches the scheduler
thread to do the computation. The scheduler thread is assigned with label
\texttt{L\_sched} and ownership \texttt{O\_sched}. 
Note that the labels/ownerships are typically programmed as variables, not
constants.  These variables are then instantiated at run time to specific
values (e.g., according to a config file).  In sum, these annotations will let
\textsc{Arbiter} enforce the separation of privileges shown in Table 1.
It should be pointed out that \textsc{Arbiter} is focused on data-centric
privilege. Thus, OS-level mechanisms like mandatory access control (e.g.
SELinux \cite{SELinux}), capability (e.g. Capsicum \cite{capsicum}), or DIFC
(e.g. Flume \cite{Flume}) can be utilized to provide complementary privilege
separation.
{\small
\linespread{0.7}
\begin{lstlisting}
dispatcher() {
  // dispatch worker to serve request
  O_user[0] = {ar, aw}; 
  O_user[1] = {br, bw}; 
  ab_pthread_create(thrW[i], NULL, thr_worker, argW[i],{dr},O_user[i]);
  ...
  // dispatch scheduler to compute 
  L_sched = {ar, br};
  O_sched = {ar, br, dr, dw};
  ab_pthread_create(thrS, NULL, thr_scheduler,argS,L_sched,O_sched);
  ...
}
\end{lstlisting}
}

As shown above, fine-grained access control inside multithreaded
  applications is achieved in a simple manner. 
  Programmers' effort is thus minimized. They can explicitly express the 
  desired security policy while still preserving the multithreaded programming
  paradigm. This property is primarily enabled by the \textsc{Arbiter} API.

\vspace{-4pt}
\subsection{A{\secfntab RBITER} API}

\textsc{Arbiter} API has three subcategories which are used for labeling,
threading, and sharing. Figure 1 lists a representative part of them.  With
these API calls, programmers can create threads that will be granted with
certain labels/ownerships at run time.  They can also have dynamic data
allocated with particular labels.

\begin{figure}[t]
\centering
\fbox{
\parbox{300pt}{
{\small
\begin{itemize}\itemsep1pt \parskip0pt \parsep0pt
\linespread{0.5}
\setlength{\leftskip}{-10pt}
\vspace{-5pt}

\item \texttt{cat\_t create\_category(cat\_type t);}

Create a new category of type \texttt{t}, which can be either 
secrecy category \texttt{CAT\_S} or integrity category \texttt{CAT\_I}.

\item \texttt{void get\_label(label\_t L);}

Get the label of a thread itself into $L$.

\item \texttt{void get\_ownership(own\_t O);}

Get the ownership of a thread itself into $O$.

\item \texttt{void get\_mem\_label(void *ptr, label\_t L);}

Get the label of a data object into $L$.
\item \texttt{int ab\_pthread\_create(pthread\_t *thread, 
	const pthread\_attr\_t 
	*attr, void *(*start\_routine)(void *),\linebreak 
	void *arg, label\_t L, own\_t O);}

Create a new thread with label $L$ and ownership $O$.

\item \texttt{int ab\_pthread\_join(pthread\_t thread, void **value\_ptr);}

Wait for thread termination.

\item \texttt{pthread\_t ab\_pthread\_self(void);}

Get the calling thread ID.

\item \texttt{void *ab\_malloc(size\_t size, label\_t L);}

Allocate dynamic memory on ASMS with label $L$.

\item \texttt{void ab\_free(void *ptr);}

Free dynamic memory on channel heap.

\item \texttt{void *ab\_calloc(size\_t nmemb, size\_t size, label\_t L);}

Allocate an array of memory on channel heap with label $L$.

\item \texttt{void *ab\_realloc(void *ptr, size\_t size);}

Change the size of the memory on channel heap.

\item \texttt{void *ab\_mmap(void *addr, size\_t length, int prot,
	int flags, int fd, off\_t offset, label\_t L);}

Map files on ASMS with label $L$.
\vspace{-5pt}
\end{itemize}
} 
} 
} 
\vspace{-5pt}
\caption{A partial list of A{\small RBITER} API}
\end{figure}

To ease the programmers' work of either porting existing programs or writing 
new ones, \textsc{Arbiter}'s memory allocation and thread library are 
fully compatible with the C Standard Library and the \textsf{Pthreads} Library. 
First, their syntax are in exact accordance. For example, \textsf{glibc} 
\texttt{malloc} also takes a \texttt{size} argument of \texttt{size\_t}
type and returns a \texttt{void} pointer pointed to the 
allocated memory chunk. 
Second, if programmers use \texttt{ab\_malloc} without 
assigning any label (\texttt{L = NULL;}), it will behave in the same way 
as \textsf{glibc} \texttt{malloc}, 
i.e., allocating a memory chunk read-writable to every thread. 
This makes it possible that programmers can 
incrementally modify their programs to run on \textsc{Arbiter}.

\vspace{-2pt}
\subsection{Security Model}
To realize data-level privilege separation, we need a set of labels for
programmers to properly express security policy. We adopted the notion used in
HiStar's DIFC label model \cite{HiStar}.  DIFC's label is a better fit for
\textsc{Arbiter} than per-data ACLs in that it is also data-centric and can
set privileges for principals in a decentralized manner.  However, it should
be noted that although \textsc{Arbiter} uses DIFC label notions, itself is not
a DIFC system (see Section 8 for more discussion).  Therefore,
\textsc{Arbiter} does not perform information flow tracking or taint
operations.

In light of the HiStar's notation, we use \emph{label} to describe the
security property of principals (i.e. threads) and ASMS data objects.  A
\emph{label} $L$ is a set that consists of \emph{secrecy} and/or
\emph{integrity categories}.  For a data object, secrecy categories and
integrity categories help to protect its secrecy and integrity, respectively.  
For a thread, the possession of a secrecy category ($*_{r}$, where $*$
represents the name of a category) denotes its \texttt{read} permission to
data objects protected by that category; likewise, an integrity category
($*_{w}$) grants the thread with the corresponding \texttt{write} permission.
We also use the notion \emph{ownership} $O$ to mark a thread's privilege to
bypass security check on specific categories.  A thread that creates a
category also \emph{owns} that category.  Different from threads, data objects
do not have ownership.  Note that categories can be dynamically created by
threads; but labels, once assigned, are not allowed to be changed.

The security in \textsc{Arbiter} is guaranteed by the following three 
rules:

\textbf{RULE 1 - Data Flow} We use $L_{A} \sqsubseteq L_{B}$, 
to denote that data can flow from $A$ to $B$ ($A$ and $B$ represent 
threads or data objects). This means: 
1) every secrecy category in $A$ is present in $B$; and 
2) every integrity category in $B$ is present in $A$. 
If the bypassing power of ownership is considered, 
a thread $T$ can read object $A$ iff 
\begin{displaymath}
	L_{A}-O_{T} \sqsubseteq L_{T}-O_{T},
\end{displaymath} 
which can be written as 
\begin{displaymath}
	L_{A} \sqsubseteq_{O_{T}}L_{T}.
\end{displaymath} 
Similarly, thread $T$ can write object $A$ iff
\begin{displaymath}
	L_{T} \sqsubseteq_{O_{T}}L_{A}.
\end{displaymath} 

In Table 1, for example, Alice's calender has label $\{d_{r}, a_{r}, a_{w}\}$.
Bob has label $\{d_{r}\}$ and ownership $\{b_{r},b_{w}\}$.
  The relation between them is that
  $L_{Alice's Cal} \not\sqsubseteq_{O_{Bob}}L_{Bob}$ and 
  $L_{Bob} \not\sqsubseteq_{O_{Bob}}L_{Alice's Cal}$. Therefore, Bob does
  not have either read or write access to Alice's calender.
  However, since the scheduler has ownership $\{a_{r},b_{r},d_{r},d_{w}\}$ and thus
  the relation $L_{Alice's Cal} \\\sqsubseteq_{O_{sched}}L_{sched}$, 
  the scheduler can read Alice's calender. Still, the scheduler 
  cannot corrupt Alice's calender because
  $L_{sched} \\\not\sqsubseteq_{O_{sched}}L_{Alice's Cal}$.
In the same way, the whole table of access rights can be figured out and,
  thus, the security concerns SC1, SC2, and SC3 can be addressed.
Meanwhile, the labeling semantics are easy to understand. For example,
  the label of Alice's calendar $\{d_{r}, a_{r}, a_{w}\}$ simply implies
  that it is readable by the scheduler and read-writable by Alice.

\textbf{RULE 2 - Thread Creation} Thread creation is another way of data
flow and privilege inheritance. Therefore, a thread $T$ is allowed to create 
a new thread with label $L$ and ownership $O$ iff 
\begin{displaymath}
	L_{T}\sqsubseteq_{O_{T}}L, O\subseteq O_{T}.
\end{displaymath}

Of course, the new thread is not allowed to modify its own labels.

\textbf{RULE 3 - Memory Allocation} Memory allocation also implies that 
data flows from a thread to the allocated memory. As the result, 
a thread $T$ can get a memory object allocated on ASMS with label $L$ iff 
\begin{displaymath}
	L_{T} \sqsubseteq_{O_{T}} L.
\end{displaymath}

With the above rules, security critical operations can be mediated and regulated.
The \textsc{Arbiter} will, based upon the assigned labels/ownerships and the 
  above rules, infer, configure, and enforce the corresponding security policy.

\section{Design}
\vspace{-5pt}
\subsection{Design Goals}
The following bullets summarize the three major goals that guided our design. 
\vspace{-6pt}
\begin{itemize}\itemsep0pt \parskip0pt \parsep0pt
\setlength{\leftskip}{-10pt}
	\item Data-centric (orthogonal to OS-level) privilege 
		separation in multithreaded applications (A1).
	\item Secure inter-thread communication (B1) 
		with fine-grained access control at data-object level (B2).
	\item To preserve multithreaded programming paradigm (C1) 
		with a unified model (rather than per-application security 
		protocols) (C2) for programmers 
		to explicitly control data flow inside applications (C3).
\end{itemize}
\vspace{-6pt}

To realize A1, we designed a novel framework for multithreaded applications
  (Section 3.2). 
To address B1 and C1, we created a new abstraction named \textsc{Arbiter} 
  Secure Memory Segment (Section 3.3). 
To tackle B2, C2, and C3, we designed \textsc{Arbiter} library (Section 3.4).
Figure 2 sketches the architecture of \textsc{Arbiter} system.

\begin{figure}[tbp]
\vspace{-12pt}
\centering
\epsfig{file=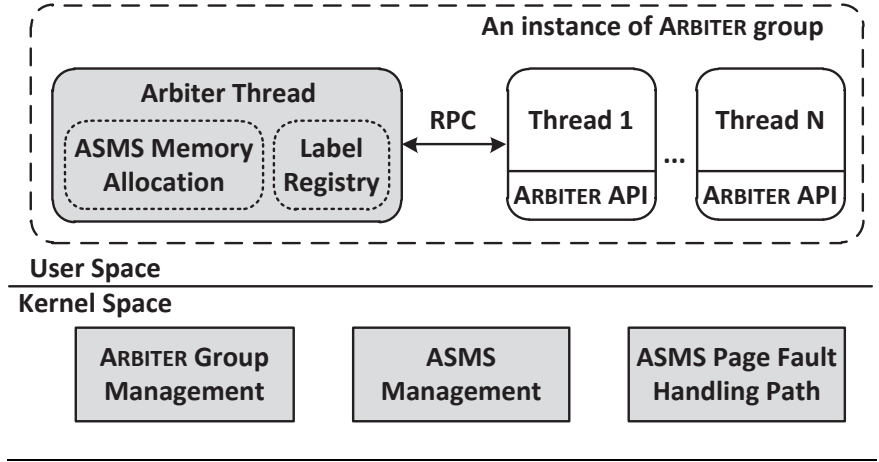, width=0.7\textwidth}
\vspace{-10pt}
\rule{0.7\textwidth}{.5pt}
\caption{System architecture}
\end{figure}

\subsection{A{\secfntab RBITER} Framework}
In \textsc{Arbiter} framework, a multithreaded program runs inside a special
  thread group, called \textsc{Arbiter} group. In addition to the application
  threads or what we call \emph{member threads}, there is a privileged thread: 
  \emph{arbiter thread}.
Arbiter thread is essentially a reference monitor, isolated from member threads. 
Its privilege does not refer to root privilege or privilege of 
  OS-level operations. 
  Instead, it means the privilege to manage the \textsc{Arbiter} group, for 
  example, to manage the label/ownership information for the 
  member threads.
The member threads, on the other hand, are the application threads being 
  protected. They run on top of the \textsc{Arbiter} API.
For those common processes and threads running on the same OS, they are not
  influenced by the existence of \textsc{Arbiter} group.
Meanwhile, multiple \textsc{Arbiter} groups can co-exist on the same OS
  without disturbing each other. This means that many \textsc{Arbiter}
  protected applications can concurrently execute.

\textsc{Arbiter} achieves privilege separation by applying address
  space isolation to the threads.
  However, it is significantly different from the multiprocessed programming: 
  the member threads still have the luxury of direct sharing of resources. 
  For example, they share code, global data, open files, signal handlers, etc.
  This is achieved by mapping the corresponding virtual memory segments to the
  same set of physical memory.
In order to ensure security, the privilege of member threads to 
  access shared resource is different from that in a traditional 
  multithreaded program. For example, stack segment is made private to each 
  member thread and inaccessible to others so that stack integrity can be 
  preserved. For similar reasons, the heap is also made private.
  Now the remaining challenge is how to allow legitimate data objects to 
  be shared efficiently and securely between member threads. 
  To tackle this problem, we created a new memory segment called 
  \textsc{Arbiter} Secure Memory Segment.

\subsection{A{\secfntab RBITER} Secure Memory Segment (ASMS)}
To preserve the merits of multithreaded programming paradigm, such as
  reference passing and data sharing, \textsc{Arbiter} needs to support
  legitimate communication between member threads. 
What makes this even more challenging is that the communication should be
  secure.

Existing approaches seem to have difficulties to achieve this goal. 
First, message passing scheme is a preferable solution for 
  program partitioning and multi-process architecture.
  However, message passing is considered as less efficient 
  than direct memory sharing \cite{Singularity:Language}, 
  especially when security is concerned so that 
  a complex message passing protocol is needed. Anyway, it dramatically changes the
  multithreaded programming paradigm which unfortunately increases the burden 
  of the program developer.
Second, some OS utilities can help to realize direct memory sharing, for example,
  Unix system call \texttt{mmap} and System V IPC \texttt{shmget}.
  However, none of them can satisfy the complex security needs in multi-principal 
  situations, such as those in our motivating example. The reason is that 
  the owner cannot directly assign different access rights for different
  principals.
Third, file system access control mechanisms, such as Unix world/group/owner
  model, can achieve the security requirements in multi-principal situations. 
  Nevertheless, it has to assign a unique UID for each thread and, even worse,
  change the programming paradigm by relying on files to share data. 

\begin{figure}[t]
      \begin{center}
          \epsfig{figure=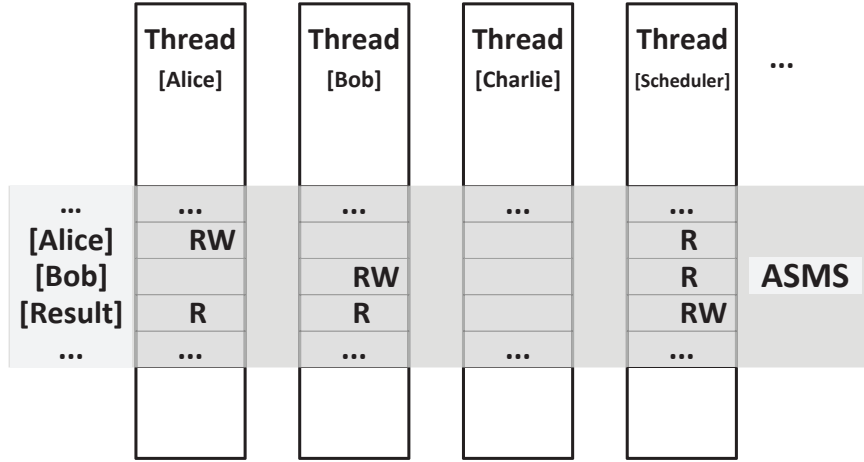, width=0.7\textwidth}
      \end{center}
      \vspace{-10pt}
      \caption{ASMS setup in the motivation example}
      \label{fig:calendar}
  \end{figure}

We created a novel memory segment called \textsc{Arbiter} Secure Memory Segment 
  (ASMS) to achieve secure and efficient sharing.
ASMS is a special memory segment compared with other memory segments like code, 
  data, stack, heap, etc. In kernel, the corresponding memory regions of ASMS
  is of a special type and thus cannot be merged with other types of memory
  regions. Regarding virtual memory mapping style, it is also different
  from both anonymous mapping and file mapping. It has a 
  synchronized virtual-to-physical memory mapping for all the member threads 
  inside an \textsc{Arbiter} group, yet the access permissions could be different. 
  Figure \ref{fig:calendar} illustrates the ASMS setup for the calendar
  example described in Section 2.1.
  We leverage the paging scheme to achieve this property.
  Furthermore, only the arbiter thread has the privilege to control ASMS. This means
  that the member threads delegate the memory allocation/deallocation requests as well
  as access rights configurations of ASMS to the arbiter thread. 
  Member threads, on the other hand, cannot directly allocate/deallocate memory on ASMS.
  Neither can they modify their access rights to ASMS on their own.

The page fault exception handling routine for ASMS is also different from 
  common page fault handling, such as copy-on-write and demand 
  paging. Legal page faults on ASMS will trigger a particular handler similar
  to demand paging. The shared page frame rather than a free frame will be mapped.
  For illegal page faults, which are mostly security violations in our case,
  either a segmentation fault by default or a programmer-specified handling routine 
  will be triggered.

Since ASMS is a kernel abstraction, it exports primitives to the arbiter thread to 
  conduct operations on it. These operations include memory allocation 
  and deallocation, access rights configuration, and so on.

\subsection{A{\secfntab RBITER} Library}
\textsc{Arbiter} library is a user-space run-time library built on top of 
  \textsc{Arbiter}'s kernel primitives. It strives to provide a fine-grained
  secure sharing mechanism based on ASMS as well as a unified model for 
  programmers to explicitly control data sharing.
  In terms of component, it has a back end (i.e. the arbiter thread), a front end
  (i.e. the \textsc{Arbiter} API), and a remote procedure call (RPC) connection 
  in between. 

The real difficulty here is how to achieve fine-grained access 
  control for program data objects on ASMS. 
According to the paging scheme and page table construct, access rights of 
  a page are described by a corresponding page table entry. Therefore, 
  traditional memory allocation mechanisms, such as \textsf{dlmalloc} 
  \cite{dlmalloc}, are not suitable because all the data objects on a page 
  can only have one kind of access rights for a certain thread.
So an intuitive idea is to allocate one page per data object (or multiple pages
  if the data object is large enough). However, this is not practical because of 
  two reasons. First, page allocation and permission configuration usually 
  requires a system call to 
  the kernel. One data object per system call will undoubtedly induce too much 
  overhead. Second, a huge amount of internal fragmentation will be incurred if
  the size of data objects is much smaller than the page size,
  which is usually the case.

To solve this problem, we devised a special memory allocation mechanism for 
ASMS: \emph{permission-oriented} allocation. The key idea is to put data objects with 
  identical labels (thus the same access rights for a thread) on the same page. 
  Allocator allocates from the system one memory \emph{block} per time so as to 
  save the number of system calls. Here a memory block is a continuous memory
  area containing multiple pages. Figure 4 demonstrates this idea (more details
  in Section 4.2).
In this way, both the internal fragmentation and performance overhead can be
  reduced to a large extent. 

At the same time of memory allocation, the arbiter thread also needs to correctly 
  determine and configure the access rights of each ASMS block for different
  member threads. To do this, the arbiter thread maintains a real-time registry 
  of the label/ownership information for member threads and ASMS memory blocks.
For example, if there is a request to allocate a new block with a certain label, 
  the arbiter thread first looks up the 
  label/ownership information for every thread. 
  Then it determines the access permissions for 
  every member thread according to the data flow rule (RULE 1).
  After that, it calls the kernel primitives to construct memory regions and
  configure access permissions.

\begin{figure}[tbp]
\centering
\epsfig{file=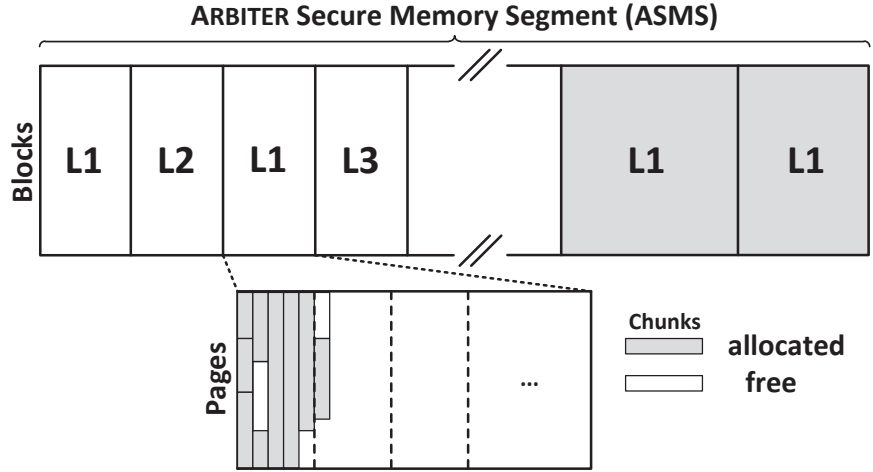, width=0.7\textwidth,}
\vspace{-10pt}
\rule{0.7\textwidth}{.5pt}
\caption{A typical memory layout of ASMS}
\vspace{-15pt}
\end{figure}

\section{Implementation}
We implemented a prototype of \textsc{Arbiter} based on Linux.  This section
highlights the major components of the implementation, including
\textsc{Arbiter} group management (Section 4.1), ASMS memory management
(Section 4.2), and page fault handling (Section 4.3).

\subsection{Management of A{\secfntab RBITER} Group}
\noindent
\textbf{Application Startup } 
Each \textsc{Arbiter} group has the preliminary environment settings that
  need to be done before the application runs. Thus, the arbiter thread 
  executes first and acts as the bootstrap of the protected application.
Arbiter first registers its identity to the kernel in order to get privileges
  to perform subsequent operations on the \textsc{Arbiter} group and ASMS 
  management. We implemented a system call \texttt{ab\_register} for this purpose.
Next, the arbiter thread starts the application by using the combination of Unix operations
  \texttt{fork} and \texttt{exec} with provided arguments,
  and blocks until a request coming from member threads.

Once the application is started, it can create child member thread by calling
\texttt{ab\_pthread\_create} and \texttt{ab\_pthread\_join} to wait for
threads' termination.  We implemented it using system call \texttt{clone}.
Registration to kernel, also via \texttt{ab\_register}, is conducted before
the new thread starts to execute.  The label and ownership of the new thread,
if not specified, default to its parent's.  \texttt{ab\_pthread\_join} is
implemented using \texttt{waitpid}.  Although implemented differently from
\textsf{Pthreads} library, we have managed to offer a compatible syntax and
functionality. 

\vspace{5pt} \noindent
\textbf{A{\small RBITER} Group Identification  } 
In order to determine privileges for the threads in an \textsc{Arbiter} group, 
  OS kernel needs to have an identification mechanism to distinguish: 
  (1) member threads from the arbiter thread, 
  (2) one \textsc{Arbiter} group from another, and 
  (3) threads in \textsc{Arbiter} groups from non-member threads, 
  i.e., normal processes and threads. 
We utilized the process descriptor structure \texttt{task\_struct} to store
  the identity information.
Specifically, we add two fields to the process descriptor. 
One is \texttt{ab\_identity}, a 32-bit identification 
  whose upper 31 bits stores the \textsc{Arbiter} group ID 
  and the last 1 bit indicates the role of either arbiter or 
  member threads.
All the threads in an \textsc{Arbiter} group share 
  the same group ID. This means that a total number of $2^{31}$ \textsc{Arbiter}
  groups can concurrently exist, which is sufficient for a general purpose OS.
Note that \texttt{ab\_identity} for a normal process is 0. 
The other item we added is \texttt{ab\_task\_list}, a doubly-linked list connecting
  every member in an \textsc{Arbiter} group, including the arbiter thread. 
Therefore, given a member thread, kernel procedures are able to 
  find its arbiter thread and peer threads.

\vspace{5pt} \noindent
\textbf{Label Registry  }
Another important aspect of \textsc{Arbiter} group management is to keep track
  of labels/ownerships of member threads and ASMS data objects.
  Whenever a new thread is created or a new data object is allocated, 
  arbiter thread should store their label/ownership information. Meanwhile, when
  to make a decision based on RULE 1, 2, and 3, arbiter thread must also lookup
  the label/ownership efficiently.
As a result, we implemented a registry and a set of storing/retrieving facilities. 
  The registry are essentially two hash tables, one is for member threads
  and other for data objects. For the sake of efficiency and simplicity, 
  the latter one is combined with the hash table for memory allocation.

\vspace{5pt} \noindent
\textbf{RPC  } 
The arbiter thread plays a major role in \textsc{Arbiter} group management by
  performing privileged operations, such as allocating ASMS data objects 
  for member threads.
Therefore, a reliable and efficient RPC connection between member threads
  and the arbiter thread is quite necessary. 
We selected Unix socket to implement the RPC.
The major advantage of Unix socket in our context is about security. 
  Unix sockets allow a receiver to get the Unix credentials, such as PID,
  of the sender. Therefore, the arbiter thread is able to verify the identity 
  of the sender thread. 
This is especially meaningful in case that
  the sender thread is compromised and manipulated by the attacker
  to send illegal requests or forged information to the arbiter thread.

\vspace{5pt} \noindent
\textbf{Authentication and Authorization  }
Every time the arbiter thread is about to handle a RPC, it first performs
  security verifications, including authentication and authorization.
Authentication helps to make sure the caller is a valid member
  thread in that group. The arbiter thread authenticates a caller by verifying the
  validity of its PID acquired from the socket. 
Authorization ensures that the caller has the 
  privilege for the requested operation, for example, to create a child 
  thread or to allocate an ASMS data object. 
  This is achieved by label/ownership comparison according to the RULE 2 and 
  RULE 3. 
If either one of the above verifications fails, the arbiter thread simply returns 
  the RPC with an indication of security violation. 
Otherwise, arbiter thread continues to perform the rest of 
  the handling procedure.

\subsection{ASMS Memory Management}
\noindent
\textbf{ASMS Memory Region Handling  }
ASMS is a special memory segment compared with other segments like
  code, data, stack, and heap. Therefore, the kernel-level memory region
  descriptors for ASMS, i.e., structure \texttt{vm\_area\_struct} in Linux, 
  should also be different from those of others. 
To do this, we added a special flag \texttt{AB\_VMA} to the \texttt{vm\_flags}
  field of the \texttt{vm\_area\_struct} descriptor.
We also want to make sure that only the arbiter thread can do allocation,
  deallocation, and protection modification of ASMS memory regions.
So we modified related system calls, such as \texttt{mmap} and \texttt{mprotect},
  to prevent them from manipulating ASMS.  

To properly create or destroy ASMS memory regions so as to enlarge or shrink
  ASMS, we implemented a set of kernel procedures similar to their Linux 
  equivalents such as \texttt{do\_mmap} and \texttt{do\_munmap}.
The difference is that when creating or destroying ASMS memory regions for 
  a calling thread, the operation will also be propagated to all the other 
  member threads in that \textsc{Arbiter} group. 
The doubly-linked list \texttt{ab\_task\_list} helps to 
  guarantee that every member threads can be traversed.
Once these batched operations are complete, each member thread will have 
  a memory region, with the same virtual address range, created or deleted.
  Nevertheless, the page access rights described in each memory region
  could be different, based on the decision made by the arbiter thread.
In order to know the arbiter thread's decision, kernel procedures export several 
  system calls to the arbiter thread. They include \texttt{absys\_sbrk}, 
  \texttt{absys\_mmap}, and \texttt{absys\_mprotect}, which have similar
  semantics as their Linux equivalents. 

In our current implementation, ASMS has a default size limit of 500MB,
  ranging from the virtual address of \texttt{0x800000 00} to \texttt{0x9fffffff}. 
  In practice, however, both the location and the size
  of ASMS is adjustable by the programmer.

\vspace{5pt} \noindent
\textbf{User-space Memory Allocation  }
For arbiter thread to handle \texttt{ab\_malloc}/\texttt{ab\_free} requests from 
  member threads, we implemented a user-space memory allocation mechanism 
  built on top of the system calls for ASMS management.
To realize fine-grained access control of data objects on ASMS, this
  allocation mechanism is permission-oriented: to put data objects with 
  identical labels on the same page.
To reduce the frequency of system calls, a block containing multiple pages
  instead of a single page is allocated per time.

Blocks are sequentially allocated from the start of ASMS. 
Of course, some data objects might have larger size and cannot fit in a 
\emph{normal block}. In this case, \emph{large blocks} will be allocated backward 
  starting at the end of ASMS. 
The patten of this memory layout is shown in the top half of Figure 4.
Inside each block, we took advantage of the \textsf{dlmalloc} strategy 
  \cite{dlmalloc} 
  to allocate memory chunks for each data object. 
The bottom half of Figure 4 depicts the memory chunks on pages inside a block.

\vspace{5pt} \noindent
\textbf{Algorithms  }
A high-level presentation of the allocation/deallocation algorithm
  is shown in Figure 5. 
For clarity, we omit the discussion on the strategy of 
  memory chunk management, which is adopted from \textsf{dlmalloc}.

\vspace{-5pt}
\begin{itemize}\itemsep0pt \parskip0pt \parsep0pt
\setlength{\leftskip}{-12pt}
\linespread{0.9}
\item \textbf{Allocation}
If the size of the data is larger than a normal block size
  (i.e. threshold), a large block will be allocated using \texttt{absys\_mmap} (line 5).
Otherwise, the arbiter thread will search for free chunks inside blocks
  with that label (line 7). If there is an available free chunk, 
  the arbiter thread simply returns it.  If not, the allocator will allocate a 
  new block using \texttt{absys\_sbrk} (line 12).

\item \textbf{Deallocation}
For a large block, the arbiter thread simply frees it using 
  \texttt{absys\_munmap} (line 3) so that it can be reused later on. 
Otherwise, the arbiter thread 
  puts the chunk back to the free list (line 5). 
Next, the arbiter thread checks if all the chunks on this block is free. 
  If so, this block will be recycled for later use (line 7).
\end{itemize}
\vspace{-5pt}

\begin{figure}[t]
    \centering
\lstset{
	language=C,
	basicstyle=\small,
	numbers=left,
	keepspaces=true,
	frame=topline,
	xleftmargin=70pt,
	xrightmargin=60pt,
	framexleftmargin=20pt,
	morekeywords={ablib\_malloc, ablib\_free}
}
\linespread{0.4}
\begin{lstlisting}
ablib_malloc(sz, L)
    if sz > threshold
        for every member thread
            Compute permission 
	    Allocate a large block
        return
    Search free chunks in blocks with label L
    if there is an available free chunk
        return
    for every member thread
        Compute permission 
        Allocate a normal block
    return
\end{lstlisting}
\lstset{frame=bottomline}
\vspace{-5pt}
\begin{lstlisting}
ablib_free(ptr)
    if it is a large block
        Free the block 
        return
    Free the chunk pointed by ptr
    if the whole block is free now
        Free the block
    return
\end{lstlisting}
\vspace{-10pt}
\caption{ASMS memory allocation algorithm}
\vspace{-10pt}
\end{figure}

\vspace{2pt} \noindent
\textbf{ASMS Initialization for New Thread  }
In addition to memory allocation/deallocation for existing member
  threads, another important responsibility of the arbiter thread is 
  to initialize the ASMS for newly created threads.
When handling \texttt{ab\_pthread\_create},
  the arbiter thread walks through every existing block.
  For each block, it determines the access permission for the new thread by 
  comparing the labels/ownerships. Then it calls 
  \texttt{absys\_mprotect} to finish the access rights configuration. 
Therefore, when the new thread starts to execute, its ASMS has already been
  correctly set up.

\subsection{Page Fault Handling}
In our implementation of \textsc{Arbiter}, two types of page faults need 
  special treatment.
One is the page fault caused by accessing ASMS. 
The other is the page fault caused by accessing data segment, 
  including both initialized data (\texttt{.data}) and 
  uninitialized data (\texttt{.bss}). 
Figure 6 depicts the flow diagram of the page fault handler. 

\begin{figure}[tbp]
\vspace{-10pt}
\centering
\epsfig{file=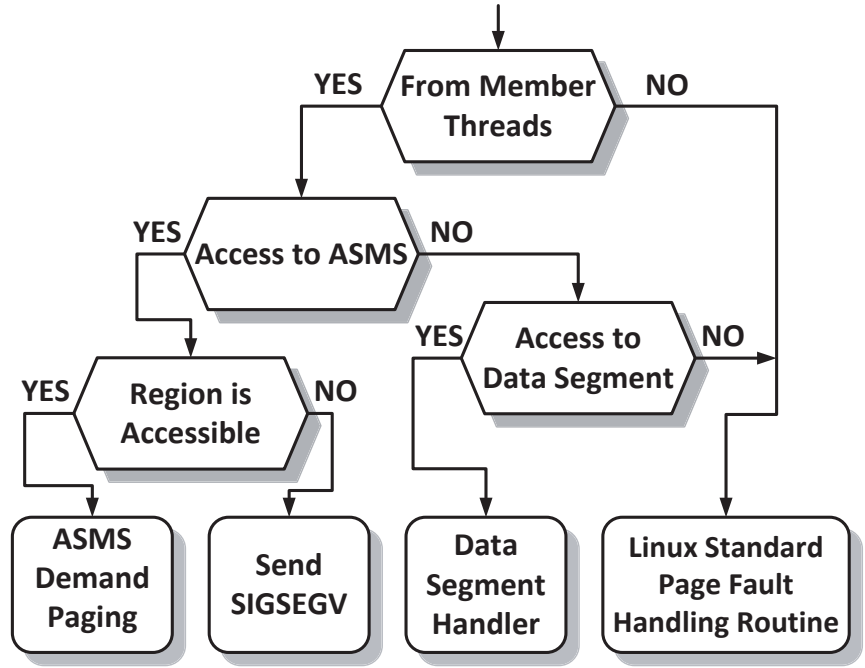, width=0.7\textwidth,}
\vspace{-10pt}
\rule{0.7\textwidth}{.5pt}
\caption{Page fault handling diagram in A{\small RBITER}}
\vspace{-15pt}
\end{figure}

\vspace{5pt} \noindent
\textbf{ASMS Fault  }
A page fault caused by accessing ASMS normally leads to two possible results:
  ASMS demand paging and segmentation fault.
ASMS demand paging happens when a member thread legally accesses an ASMS page
  for the first time. In this case, the page fault handler should
  find the shared physical page frame and create and configure the corresponding page 
  table entry for the member thread. The protection bits of the page table 
  entry are determined according to the associated memory region descriptor.
In this way, subsequent accesses to this page will be automatically checked by MMU 
  and trapped if illegal, for example, trying to write with read-only permission.
This hardware enforced security check significantly contributes to the 
  satisfying run-time performance of \textsc{Arbiter}.
Illegal access to an ASMS page will result in a \texttt{SIGSEGV} signal sent to
the faulting thread.  Then it is up to the programmer to either, by default,
let the thread terminate or customize a handling routine.
We implemented a kernel procedure \texttt{do\_ab\_page} and made corresponding
  modifications to the Linux default page fault handler to realize the 
  above idea. 

\vspace{5pt} \noindent
\textbf{Data Segment Fault  }
Most multithreaded programs also use global data to share information. 
In order to minimize programmers' porting effort, we decide to make
  the data segment accessible to all member threads
  and programmers only need to move security-critical global data
  on to ASMS instead of all of them.
If data segment is made private for each thread, programmers will be forced
  to declare all the global data on ASMS. Though technically feasible,
  it inevitably increases programmers' porting effort.
Meanwhile, our observation indicates that global data is usually not security 
  sensitive because a large portion of it is metadata rather than user data.
Therefore, we decided to make the data segment accessible to all member threads.
In this way, programmers only needs to move the security sensitive global data 
  into ASMS instead of all of them. 
We implemented another page fault handling subroutine to do this. 
  Since data segment is originally shared after \texttt{clone}, this subroutine 
  mainly helps to avoid the copy-on-write operations.
 
A minor yet critical issue with data segment page fault handling is about 
  the futex (i.e. fast userspace mutex) mechanism.
Multithreaded programs often rely on mutexes and 
  condition variables, which are often declared on data segment, for mutual exclusion
  and synchronization. 
In \textsf{Pthreads}, both of them are implemented using futex. 
  For retrieving purpose, kernel regards the key of each futex as either 
  the address of \texttt{mm\_struct} if the futex is on an anonymous page 
  or the address of \texttt{inode} if the futex is on a file mapped page.
  In \textsc{Arbiter}, since data segment is anonymous mapping but the 
  \texttt{mm\_struct} for each member thread are different,
  kernel will treat the same mutex or 
  condition variable as different ones. Of course, we 
  can force programmers to declare them on ASMS, which is file mapping
  and does not have this issue.
  However, we again decided to reduce programmers' effort by modifying the 
  corresponding kernel routine \texttt{get\_futex\_key} and set the 
  key to a same value.

\subsection{Implementation Complexity}
We implemented \textsc{Arbiter} prototype using C. 
We add about 2,000 LOC to the Linux kernel. 
The arbiter thread including memory allocation, part of which is based on 
  \textsf{uClibc} \cite{website:uClibc}, contains around 3,500 LOC.
The \textsc{Arbiter} API contains roughly 600 LOC. 
\textsc{Arbiter}'s trusted computing base (TCB) consists of the arbiter thread and
  the kernel, as shaded in Figure 2. 
  The \textsc{Arbiter} API is not part of the TCB. 

\section{Application}
To explore \textsc{Arbiter}'s practicality on ease of adoption and 
  to examine its effectiveness to enhance application security,
  we ported \textsf{memcached} \cite{website:memcached} to \textsc{Arbiter}
  system.
Section 5.1 introduces \textsf{memcached}.
Section 5.2 presents the effort we applied to retrofit \textsf{memcached}.
Section 5.3 discusses how \textsf{memcached} could benefit from 
  \textsc{Arbiter} in terms of security. 

\subsection{Memcached}
\textsf{Memcached} is a distributed, in-memory object caching system, 
  intended for speeding up dynamic web applications by alleviating
  database load.
It caches data and objects from results of database queries, API calls, or page
  rendering into memory (i.e. RAM) so that the average response time can be reduced
  to a large extent.
It is widely-used for data-intensive websites
  including LiveJournal, Wikipedia, Facebook, YouTube, Twitter, etc.

\textsf{Memcached} uses a client-server architecture. The servers maintain
  a key/value associative array. There can be multiple clients talking to the same server.
  Clients use client libraries 
  to contact the servers to make caches or queries. 

Security issues arise considering the fact that \textsf{memcached} 
  servers are multithreaded. Two worker threads in the same server might work
  on behalf of different applications or users, which may not be mutually trusted.
  Although \textsf{memcached} has SASL (Simple Authentication and Security Layer) 
  authentication support, it cannot prevent authenticated clients from 
  querying or updating others' data, either accidentally or maliciously.
What might be even worse is that the vulnerabilities in \textsf{memcached} 
  \cite{memcached:vul, memcached:vul_more} could be exploited by an adversary
  (e.g., through a buffer overflow attack) so that the compromised worker thread can
  arbitrarily access any data on that server.
Approaches like SELinux can help to restrict OS-level operations of the compromised
  thread, but cannot deal with program data objects.

\subsection{Retrofitting Effort}
To mitigate the security problems stated above, we retrofitted a \textsf{memcached}
  server on top of \textsc{Arbiter} system.
There are mainly three types of threads in a \textsf{memcached} process:
  main thread, worker thread, and maintenance thread.
Upon arrival of each client request, the main thread 
  dispatches a worker thread to serve that request.
Periodically, maintenance threads wake up to maintain some important assets such as
  the hash table.
For demonstration clarity, we only focus on the main thread and worker threads 
  and regard the maintenance threads with the same privilege as the main thread.

To separate the data-centric privileges for each thread, we replace the original
  thread creation function \texttt{pthread\_create} with \texttt{ab\_pthread\_create}.
Regarding dynamic data that needs to be shared, we replace the 
  \texttt{malloc/free} function families with \texttt{ab\_malloc/ab\_free} 
  to allocate them on ASMS.
For those dynamic data of private use for each thread, we do not make any change
  so that they will be allocated on their own heap that is private.
Since global data are shared in \textsc{Arbiter}, we only need to declare those
  which contain sensitive information onto ASMS. 
In case of \textsf{memcached}, however, we do not find any type of this global data.

We assign each thread and data with different labels and ownerships.
For example, consider a \textsf{memcached} server which is supposed to serve two 
  applications App A and App B, or two users User A and User B.
To make the scenario more complex, we assume that A has a higher privilege to
  read (but not write) B's data.
Meanwhile, some metadata created by the main thread in \textsf{memcached} 
  are shared with worker threads for dispatching purpose.
  For instance, the shared \texttt{CQ\_ITEM} contains information about the
  connection with the client and some request details. We want to protect
  the integrity of these critical metadata so that they are read-only to
  worker threads.  
In light of these security needs, we made the corresponding configurations 
  shown in Table 2.

\begin{table}[tbp]
\centering
{\small
\setlength{\tabcolsep}{4pt}
\begin{tabular}{c|c|c|c} \hline
    & \textbf{Main/Maintenance} & \textbf{A} & \textbf{B} \\
    \textbf{Data} & L:\{\} & L:\{mr,br\} & L:\{mr\}  \\
 & O:\{mr,mw\} & O:\{ar,aw\} & O:\{br,bw\} \\ \hline \hline
 \textbf{A's Data} & \multirow{2}{*}{\textbf{--}} & \multirow{2}{*}{\textbf{RW}} & \multirow{2}{*}{\textbf{--}} \\
L:\{ar,aw\} & & &  \\ \hline
\textbf{B's Data} & \multirow{2}{*}{\textbf{--}} & \multirow{2}{*}{\textbf{R}} & \multirow{2}{*}{\textbf{RW}} \\
L:\{br,bw\} & & &  \\ \hline
\textbf{CQ\_ITEM} & \multirow{2}{*}{\textbf{RW}} & \multirow{2}{*}{\textbf{R}} & \multirow{2}{*}{\textbf{R}} \\
L:\{mr,mw\} & & &  \\ \hline
\end{tabular}
}
\vspace{-8pt}
\caption{Sample labeling configuration for {\sf memcached}}
\vspace{-15pt}
\end{table}

We modified the source code of \textsf{memcached} based on the above scenario. 
We slightly changed the original thread dispatching scheme so that requests
  from different principals can be delivered to the associated worker thread.
  This modification does not affect other features of \textsf{memcached}.
  The changes to the program are only around 100 lines of code out of 20k 
  in total. 
  Among these changes, label/ownership declaration
  takes 31 and function replacement for label annotation takes 46.

\subsection{Protection Effectiveness}
To demonstrate \textsc{Arbiter}'s protection effectiveness, 
we assume that an adversary has exploited a program flaw or vulnerability in
  \textsf{memcached} and thus take control of a worker thread. 
We have simulated various attempts to access other threads' data.
Some typical scenarios are as below.
\vspace{-5pt}
\begin{itemize}\itemsep0pt \parskip0pt \parsep0pt
\setlength{\leftskip}{-10pt}
	\item First, the adversary may try to read or write the data
		directly. A segmentation fault will then be triggered.
	\item The adversary may want to call \texttt{mprorect} to change
		the permission of ASMS. However, \textsc{Arbiter} forbids
		\texttt{mprotect} to operate ASMS memory region.
	\item The adversary may attempt to call \texttt{ab\_munmap} first and then
		call \texttt{ab\_mmap} to indirectly modify the permission
		to ASMS. However, since it does not have permission to access
		the data, arbiter basically will deny the \texttt{ab\_munmap}
		request during authorization.
	\item The adversary may also want to \texttt{fork} a child process
		outside of that \textsc{Arbiter} group. Still, the child process
		cannot modify the permission of ASMS. If the child process
		calls \texttt{ab\_register} to set itself as a new arbiter thread,
		it will have the privilege to manage ASMS. However, it is 
		in a new \textsc{Arbiter} group and its ASMS can no longer have 
		the same physical mapping as the old ASMS does.
	\item A more sophisticated approach is to forge a reference and
		fool an innocent worker thread to access data on behalf of the
		adversary.
		However, \textsc{Arbiter} provides an API \texttt{get\_privilege},
		that allows the innocent thread to verify if the requesting 
		thread has the necessary privileges. 

\end{itemize}
\vspace{-3pt}

Nevertheless, the adversary can starve resource by creating 
member threads or allocating ASMS memory. 
This can be properly addressed by enforcing a resource limit.

\section{Performance Evaluation}

In this section, we describe the experimental evaluation of \textsc{Arbiter}
  prototype. Our goal is to examine the system overhead (Section 6.1) and 
  to test the application level performance (Section 6.2).

All of our experiments were run on a server with Intel quad-core Xeon X3440
  2.53GHz CPU and 4GB memory. We used 32bit Ubuntu Linux (10.04.3) with
  kernel version 2.6.32. 
  Since we implemented the ASMS memory chunk allocation mechanism based on 
  \textsf{uClibc} 0.9.32, we used the same version as microbenchmarks for comparison.
We employed \textsf{glibc} 2.11.1 with the \textsf{Pthreads} library version 2.5 to
  compare with \textsc{Arbiter}'s thread library.
In addition, we built a security-enhanced \textsf{memcached} based on its version
  1.4.13 and we chose \textsf{libMemcached} 1.0.5 as the client library.

\begin{table}[t]
\centering
{\small
\begin{tabular}{l|c|c|c} \hline
    \textbf{Operation}&\textbf{Linux ($\mu$s)}&\textbf{A{\scriptsize RBITER} ($\mu$s)}&\textbf{Overhead}\\\hline
(ab\_)malloc & 4.14 & 9.09 & 2.20\\
(ab\_)free & 2.06 & 8.36 & 4.06\\
(ab\_)calloc & 4.14 & 8.41 & 2.03\\
(ab\_)realloc & 3.39 & 8.27 & 2.43\\
(ab\_)pthread\_create & 91.45\phantom{0} & 145.33\phantom{00} & 1.59\\
(ab\_)pthread\_join & 36.22\phantom{0} & 41.00\phantom{0}& 1.13\\
(ab\_)pthread\_self & 2.99 & 1.98 & 0.66\\
create\_category & -- & 7.17 & --\\
get\_mem\_label & -- & 7.66 & --\\
get\_label & -- & 7.65 & --\\
get\_ownership & -- & 7.55 & --\\
ab\_register & -- & 2.74 & --\\ \hline
ab\_null  & -- & 5.84 & --\\ 
(RPC round trip) & & & \\ \hline
(absys\_)sbrk & 0.65 & 0.76 & 1.36\\
(absys\_)mmap & 0.60 & 0.83 & 1.38\\
(absys\_)mprotect & 0.83 & 0.92 & 1.11\\ \hline
\end{tabular}
}
\caption{Microbenchmark results in Linux and A{\small RBITER}}
\vspace{-15pt}
\end{table}

\subsection{Microbenchmarks}

The runtime overhead induced by \textsc{Arbiter} mainly comes from the
  \textsc{Arbiter} API. Therefore, we selected these API calls as microbenchmarks
  to quantify the performance overhead. We ran each microbenchmark
  on Linux and \textsc{Arbiter}, and recorded the average time consumption
  of 1,000 times of repeat. 
In order to make a fair comparison, we ran the test program as
  a single-threaded application on \textsc{Arbiter}.
  Table 3 shows the results.

The major contributor of the latency of \textsc{Arbiter} API is the
  RPC round trip between member threads and the arbiter thread. The RPC round trip contains
  RPC marshaling, socket latency, etc.
To measure the overhead from RPC, we implemented a null API named \texttt{ab\_null}
  to test the RPC time usage. As shown in Table 3, a RPC round trip
  takes 5.84 $\mu$s on average. This result can help to explain
  most of the API overhead.
One exception is \texttt{ab\_register}, which is a direct system call and does
  not involve any RPC round trip.
Meanwhile, \texttt{ab\_pthread\_self} runs even faster than its Linux equivalent.
  This is due to our own design of thread management and thus
  a simpler implementation using \texttt{getpid} to return the thread ID.

In addition to the latency of RPC round trip, the system calls made by the 
  arbiter also contribute to the API overhead. 
  To further examine the system call overhead, we selected 
  \texttt{sbrk}, \texttt{mmap}, and \texttt{mprotect} to compare between
  Linux and \textsc{Arbiter}. On average, \textsc{Arbiter}'s version of these
  system calls have a 28\% overhead.

The above experiment does not take into account of the number of member threads. 
  Theoretically, since the memory allocation on ASMS for one member thread is
  also propagated to others, the time usage for memory allocation should be 
  proportional to the number of concurrent member threads.
Meanwhile, for those API calls dealing with label information such as \texttt{get\_label},
  operations are only conducted on one thread, i.e., the calling thread.
  Thus, in theory, the time usage of these API calls should remain constant despite the
  change of thread number.
The left subfigure in Figure 7 shows the time usage of \texttt{ab\_malloc} and \texttt{get\_label}
  as the number of member threads increases. The result conforms to our 
  theoretical prediction very well.
  The time increasing rate of \texttt{ab\_malloc} is roughly 5.7\% per additional thread.
  And the time variation of \texttt{get\_label} is almost 0.

In addition, the creation of member threads also involves the access rights
  reconfiguration for ASMS. The thread creation time should be proportional 
  to the size of allocated pages on ASMS.
The right subfigure in Figure 7 shows the time consumption of \texttt{ab\_pthread\_create}
  with regard to the
  allocated ASMS size. It is also in line with our expectation.

\begin{figure}[t]
\vspace{-5pt}
\centering
\epsfig{file=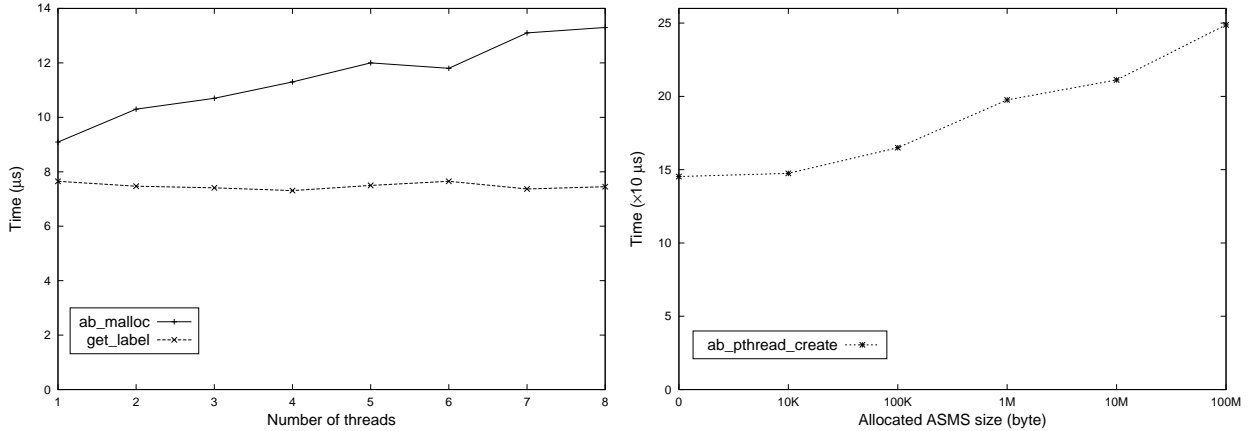, width=\textwidth,}
\vspace{-10pt}
\caption{A{\small RBITER} API performance regarding number of threads and allocated ASMS size}
\vspace{-10pt}
\end{figure}

\subsection{Application Performance}
We conducted experiments on \textsf{memcached} to further evaluate the 
  application level performance of \textsc{Arbiter}.
According to the results of the above experiments, the overhead induced by 
  \textsc{Arbiter} is related to the number of member threads.
  Hence, we stick to the scenario introduced in Section 5.2,
  in which four threads are involved
We built a client as benchmark which constantly contacts \textsf{memcached}
  on behalf of User X and User Y.
We measured the throughput of two basic \textsf{memcached} operations, 
  SET and GET, with various value size and key size.
The results are compared with that of unmodified \textsf{memcached}.
In the upper two sub of Figure 8, we anchored the key size to 32 bytes 
  and changed the value
  size. Note that we selected a non-linear distribution of value size in order 
  to cover as much range as possible.
In the lower two sub of Figure 8, we fixed the value size to 256 bytes
  and adjusted the key size.
Each point in the figure is an average of 100,000 times of repeat.
All together, the average performance decrease incurred by \textsc{Arbiter} is 
  about 5.6\%, with the worst case of 19.0\%.

Compared to the microbenchmarks, the application level
  performance overhead is relatively small.
The major reason is that the \textsc{Arbiter} specific operations (i.e. \textsc{Arbiter}
  API calls) are invoked infrequently. 
  For most of the time, \textsf{memcached} does the real work rather than
  calling \textsc{Arbiter} API.
In fact, \textsf{memcached} is a memory intensive application. For those
  applications which are not as memory intensive, we believe they could 
  achieve better overall performance.

\begin{figure}[t]
\vspace{-3pt}
\centering
\epsfig{file=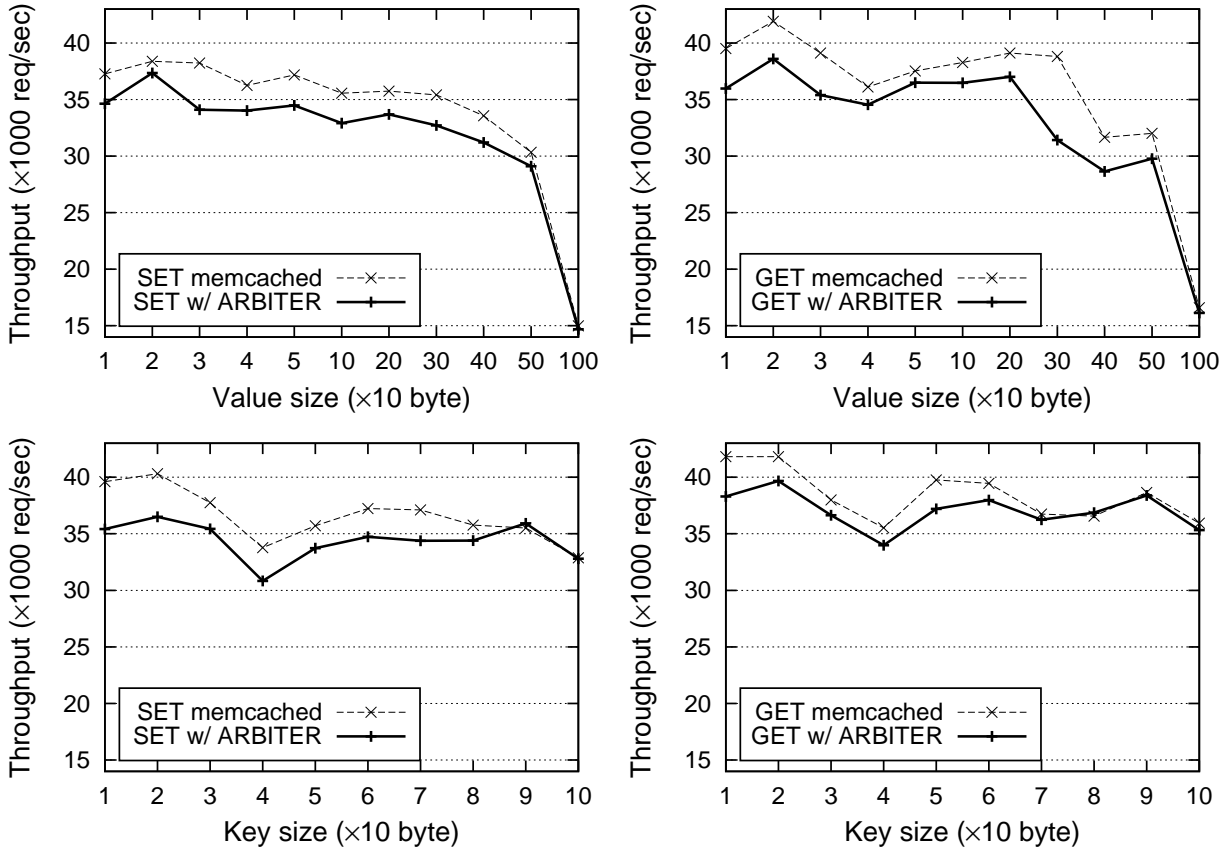, width=\textwidth,}
\vspace{-10pt}
\caption{Performance comparison for \textsf{memcached}}
\vspace{-15pt}
\end{figure}

\section{Related Work}

Privilege separation has long been the primary
principle for secure system construction. It has been
applied to system software dated from Multics to micro-kernels
\cite{sel4} and Xen \cite{xen_vee08, xoar} to build secure and resilient
operating systems and virtual machine monitors. For application
software, it has also been widely used to securely construct severs
\cite{Provos:PPE, apache, Wedge}, web applications \cite{swift_sosp07} and
browsers \cite{Chrome, gazella, op}.

A typical way to achieve privilege separation is to leverage OS-level
run-time access control techniques. For example, Apache \cite{apache}
and OpenSSH \cite{Provos:PPE} split their functionality into different
processes with different UID/GID, and leverage
OS-level access control mechanism to enforce privilege
separation. Traditionally, UNIX systems employs discretionary access
control (DAC) mechanism in which users can specify access permissions
for their own files. To further provide system-wide policies and
centralized control, mandatory access control (MAC) mechanisms like
SELinux \cite{SELinux} and Apparmor \cite{apparmor} are proposed for
commodity OSes. In MAC systems, the system administrator is
responsible for all the type assignment and policy configuration,
which could be inflexible and error-prone. Thus, other research
approaches are proposed that allow users to create protection domains
on their own. Capsicum \cite{capsicum} provides commodity UNIX systems
with practical capability primitives to support privilege
separation. Flume \cite{Flume} implements 
DIFC in Linux. Besides research effort on commodity
OSes, EROS \cite{EROS}, Asbestos \cite{Asbestos}, and
HiStar \cite{HiStar} are clean-slate designed operating
systems that take least-of-privilege as the major design principle and
offer native security primitives (capabilities or information flow
labeling) bound with OS abstractions. The primary difference between
OS-level access control systems and \textsc{Arbiter} is the type of
privilege concerned: OS-level approaches focus on managing privileges
to OS resources (e.g, files, networking, and devices) while
\textsc{Arbiter} is primarily used for controlling access privileges
of data object within a program. Thus, these approaches and
\textsc{Arbiter} are complementary to each other to build secure
programs.

Sandboxing techniques such as software fault isolation (SFI)
\cite{sfi, xfi, nacl} and binary translation \cite{vx32} are developed
to restrict the privilege of certain portion of untrusted code inside
a trusted program. The major goal of sandbox approaches is isolation
and confinement, which is to prevent untrusted code from manipulating
control and data of the main program illegally. In contrast,
\textsc{Arbiter} not only enforces isolation, but also provides
primitives for secure sharing and communication in multi-principal
programs with mutually untrusted relationships.
Wedge \cite{Wedge}, though taking heap data access into consideration, still
targets at splitting monolithic programs into functional compartments with
just-enough privileges.
\textsc{Arbiter}, however, aims to resolve the complex data sharing
relationships in multi-principal applications, which is difficult to achieve
in the traditional multithreaded programming.

Language-based approaches can help programmers enforce fine-grained
access control in regard to program internal data and
semantics. Typically these approaches integrate security notions into
the type systems of the programming language, and/or use verification
methods and compiler-inserted checks to enforce security policies and
access control. Jif \cite{JFlow} and Joe-E \cite{Joe-E} are Java
language extensions that implements DIFC and capability security
primitives, respectively. Laminar \cite{Laminar} is a Java
language-based DIFC system with the OS support to handle OS
abstractions. Singularity \cite{Singularity:Language} is a research
operating system which leverages language (C\# extension) support and
static verification to achieve isolation and controlled communication.
Aeolus \cite{Aeolus} is a DIFC platform for building secure distributed
applications using practical abstractions relying on memory-safe language.
Compared to these language-based approaches, \textsc{Arbiter} does not
require a strongly typed language such as Java and C\#. Therefore, it
takes much less effort to build or enhance the security for
multithreaded programs written in C/C++ and \textsf{Pthreads} library using
\textsc{Arbiter}.

Virtual machine offers another layer of indirection for security
mediation and enforcement \cite{split_interface}, with advantages of
transparency and tamper-proof. In \textsc{Arbiter}, we choose not to
employ virtual machine for building our security mechanisms, primarily
because of the semantic gap (e.g., managing process memory regions at the
VMM-level) and performance (e.g., system call vs. hypercall)
reasons. However, \textsc{Arbiter} could incorporate existing virtual
machine security techniques to reduce kernel TCB and improve the
security of OS kernel. These approaches include: reducing the trust of
kernel for applications \cite{overshadow, trustvisor}, kernel
integrity protection \cite{secvisor}, and sandboxing untrusted kernel
code \cite{xen_driver, huko}.

\section{Discussion}
\vspace{-5pt}
\textsc{Arbiter} is a generic approach and applicable to a variety of 
  applications. In this paper, we demonstrate its application in calender server
  and \textsf{memcached}. 
  With a fully compatible memory allocation and thread library,
  we believe that most, if not all, types of multi-principal multithreaded 
  applications can benefit from \textsc{Arbiter}. 
  Potential candidates include cloud-based collaborative applications, 
  database servers, Web servers, printing servers, etc.
  In the future, we plan to port more legacy software onto \textsc{Arbiter}
  and further demonstrate its applicability.

Nevertheless, some limitations still exist in our current design and implementation
  of \textsc{Arbiter}. 
One limitation is that the user-space ASMS memory 
  allocator uses a single lock for allocation/deallocation. Therefore,
  the processing of allocation and deallocation requests
  has to be serialized. This partly contributes to the performance overhead.
  A finer lock granularity can help to improve parallelism and scalability,
  such as the per-processor heap lock in Hoard \cite{Hoard}.
  In fact, \textsc{Arbiter}'s memory allocation mechanism inherently has the 
  potential to employ a per-label lock. This per-label lock makes \textsc{Arbiter}
  ready to embrace a much more efficient allocator. We are looking at ways
  to implement such a parallelized allocator. 

Another limitation with \textsc{Arbiter} is about the security policy
specification.
First, in complex and dynamic deployment scenarios, getting all the
labels/ownerships correct is not an easy job for system admins or programmers.
Second, in cases where programmers want to build secure-by-design applications
and only need to hard code policies, the annotations may still spread across
multiple locations, which increases the possibility of human error. 
One potential solution is to have a policy debugging or model checking tool
that can verify the correctness of label/ownership assignments.
Another direction is to let human express security requirements in a
high-level policy description language and automatically generate annotations. 
We plan to investigate both directions in the future.

\textsc{Arbiter}'s security model, including notions and rules, are inspired 
  by DIFC. However, it should be noticed that \textsc{Arbiter} does not perform 
  information flow tracking inside a program, mainly due to two observations. 
(1) For a run-time system approach, tracking fine-grained data flow (e.g. moving
  a 4-byte integer from memory to CPU register) could incur
  tremendous overhead, making \textsc{Arbiter} impractical to use. 
(2) The fact that information flow tracking can enhance security does 
  not logically exclude the possibility of solving real security
  problems without information flow tracking. 
The main contribution of \textsc{Arbiter} is that it provides 
  effective access control and practical fine-grained privilege separation
  while preserving the advantages of traditional multithreaded 
  programming paradigm. 
Once practical information flow tracking mechanisms emerge, 
  we will investigate this issue and explore the solution space 
  of integrating \textsc{Arbiter} with tracking.

\section{Conclusion}
\textsc{Arbiter} is a run-time system and a set of security primitives which
  can realize fine-grained and data-centric privilege separation for 
  multi-principal multithreaded applications.
\textsc{Arbiter} is a practical framework that not only provides
  privilege separation and access control, but also preserves the convenience and 
  efficiency of traditional multithreaded programming paradigm.
Our experiments demonstrate \textsc{Arbiter}'s ease-of-adoption
  as well as its satisfying application performance.

\bibliographystyle{abbrv}
\bibliography{arbiter}

\end{document}